\documentclass[
	a4paper, 
	10pt, 
	unnumberedsections, 
	twoside, 
]{LTJournalArticle}
\usepackage[inkscapelatex=false]{svg}
\usepackage[pdf]{graphviz}
\usepackage{xpatch}
\makeatletter
\newcommand*{\addFileDependency}[1]{
  \typeout{(#1)}
  \@addtofilelist{#1}
  \IfFileExists{#1}{}{\typeout{No file #1.}}
}
\makeatother
\xpretocmd{\digraph}{\addFileDependency{#2.dot}}{}{}
\usepackage{subcaption}

\runninghead{} 

\footertext{\textit{RISC-V Summit Europe, Paris, 12-15th May 2025}} 

\title{Modular SAIL: dream or reality?} 

\author{%
 Petr Kourzanov\textsuperscript{1}\thanks{Corresponding author: \href{mailto:peter.kourzanov@imec.be}{\tt peter.kourzanov@imec.be}}
 \and
 Anmol Anmol\textsuperscript{1}
}

\date{
\vspace{-1em}
\footnotesize\textsuperscript{\textbf{1}}imec DSRD CSA, {\em Kapeldreef 75, 3001 Leuven, Belgium}
\vspace{-1em}
}

\renewcommand{\maketitlehookd}{%
\begin{abstract}
\noindent In order to truly benefit from RISC-V ISA modularity, the community has to address the issue of compositionality, going beyond modules at the specification level covering larger subsets of the RISC-V development flow including emulation, simulation and verification.
In this paper we introduce \emph{modular SAIL}, an experiment to inject compositionality into the \emph{SAIL-RISCV golden model}. We show that it is, in principle, not difficult to adapt the SAIL-RISCV flow (and ideally the \emph{SAIL compiler} itself) to support modules at the emulator level. We back our findings by a comparative study of the resulting pluggable emulator's performance using both static and dynamic binding, which both exhibit same functional behaviour as the original monolithic emulator (aka RISC-V \textbf{ISS}).
\end{abstract}
\vspace{-1.5em}
}

\begin{document}

\maketitle 

\section{Introduction and related work}
\vspace{-.5em}

The benefits of open ISAs (e.g., RISC-V, \textbf{OpenPOWER}) and free CPUs (e.g. \textbf{OpenSPARC}) are well known \cite{RISCV}. RISC-V, on paper, has an additional bonus of offering a \emph{modular} ISA: the implementers, tool providers as well as their users can freely agree on the set of \emph{extensions} that need to be cast to HW and supported by SW. With the advent of co-processor interfaces such as OVI \cite{OVI} and XIF \cite{CV-X-IF}, an \emph{ecosystem} of extension IP vendors becomes imaginable. From standardization perspective, initiatives like CX \cite{CX} do promote this path, by pushing the interface to the ISA level.

Many of these alternatives come with different trade-offs. This can be for example in the level of SW/tooling support they require. Or the necessity of low-level assembly programming with intrinsics vs. the need to adjust the microarchitecture of the core's pipeline. Nevertheless they do share one crucial requirement - modularity of \emph{spec}, \emph{design}, \emph{testing} and \emph{verification}.

In order to truly enable such an ecosystem, the RISC-V community would address the \emph{compositionality} not only of the RISC-V specification itself, but also of all ensuing artefacts: (1) emulators, (2) simulators, (3) test-benches, (4) compiler extensions, (5) the OS and (6) runtime libraries and middleware. While the ISA itself is open, it is unlikely that many players in the eco-system will opt for open-sourcing their actual extension IP designs. With the rapid increase of the number of extensions for various kinds of accelerators for RVV, ML/AI and Crypto (and well as the number of their HW tapeouts) the modularity and compositionality of RISC-V extensions is expected to become more and more critical to the success of RISC-V community as a whole.
In this paper we focus on the first and the most crucial artefact: the SAIL-generated \verb|riscv_sim_RV64| emulator \cite{SAIL}. As SAIL-RISCV \cite{SAIL-RISCV} is selected for the role of the \emph{golden model}, it provides the ultimate source of truth with which further artefacts are measured. This includes, for example, the Spike simulator \cite{Spike}, \textbf{RVFI} tools, \textbf{gcc} \& \textbf{LLVM} compiler toolchains.

By modularizing the \verb|riscv_sim_RV64| we aim for:
\begin{enumerate}
\item minimizing cross-module dependencies
\item binary plug-ability of extension \emph{modules}
\item dynamic binding of the RISC-V ISA extensions
\item wide coverage of important \textbf{HPC} extensions: \textbf{M}, \textbf{V}, \textbf{P} (draft), \textbf{B} and \textbf{Crypto} (\textbf{K} in this paper) 
\item minimal changes to the compiler and golden model
\end{enumerate}


CBI \cite{CBI} provides a fine-grained support for introducing individual instructions to a range of their cores using a proprietary Studio Fusion tool. New FUs introduced are isolated from the base core pipeline, however, the behaviour needs to be described to the toolchain at a high-level which then generates a full RTL for the complete core (as well as the required SW tools).
OVI \cite{OVI} from \textbf{BSC} and Semidynamics offers a custom signaling interface that allows a base core to offload some specific instructions to an IP coming from a separate entity.
XIF \cite{CV-X-IF} from \textbf{OpenHW} follows a more compositional approach whereby a base core decoder filters instruction stream and forwards unknown opcodes to a co-processor which then interfaces with the LSU and the RF. 
To our knowledge, in all three approaches the verification needed before and after RTL synthesis is still performed on the whole design.
\vspace{-1em}
\section{Methodology}
\vspace{-2em}

{
\begin{figure}[h]
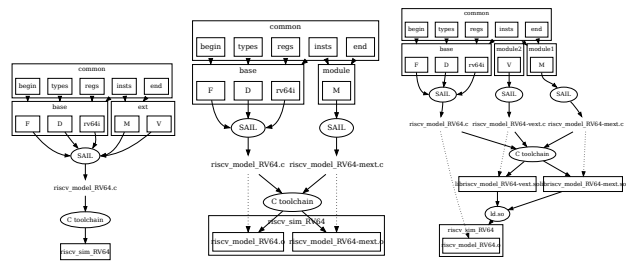

\begin{subfigure}[t]{.28\columnwidth}
\includesvg[width=\textwidth]{old.svg}
\caption{Original}
\label{fig:orig}
\end{subfigure}
\begin{subfigure}[t]{.33\columnwidth}
\includesvg[width=\textwidth]{new.svg}
\caption{Static \textbf{M}}
\label{fig:static}
\end{subfigure}
\begin{subfigure}[t]{.37\columnwidth}
\includesvg[width=\textwidth]{dyn.svg}
\caption{Semi-Dyn \textbf{M}\&\textbf{V}}
\label{fig:dynamic}
\end{subfigure}
\caption{Proposed changes to the SAIL flow}
\label{fig:proposal}
\end{figure}
}

In our experiment, most significant changes were introduced in the \verb|Makefile| of the model. As can be seen in Fig.\ref{fig:proposal}, instead of producing a single whole-program-processed emulator Fig.\ref{fig:orig}, we are segregating large extensions from each other (and from most of the base core instructions and architectural state) in Fig.\ref{fig:static}. Once this was found to be working, we enabled dynamic binding as in Fig.\ref{fig:dynamic}. For our aims, in this experiment, we are post-processing the generated C emulator(s) for each extension module automatically by a custom script, addressing the compositionality aspects with the help of the following transformations:
\begin{itemize}
\item prefixing and \emph{externalizing} symbols of the module API: \verb|zdecode|, \verb|zprint_insn|, \verb|zexecute|, \verb|{CREATE,KILL}(zast)| and \verb|model_{init,fini}|
\item prefixing and relocation of data-structure declarations from the generated C file to a H file
\item making extern the (shared) architectural state, as most registers are in the base for our experiment
\item making \emph{static} incidental definitions and helpers which might get duplicated by current approach
\item removal of (de)initialization of the (shared) exception handling and run-time system mechanisms 
\end{itemize}
    
\vspace{-2em}
\section{Implementation and Results}
\vspace{-.5em}

In effect, each module obtains a well-defined API for performing the decode-print-execute loop and for committing the results to the RF. While each generated C emulator has all the features to run just the instructions provided by the given ISA descriptions, it obviously lacks the base instructions and hence can not be run in isolation from the base emulator - which provides the system context, load/store, exception and architectural state management such as CSRs.

To integrate these extensions back into the base
emulator, we have applied a simple patch
that adds required calls to respective functions in \verb|zstep| and \verb|model_init| of the base emulator.\footnote{latest model requires a patch to \texttt{zextensionEnabled}, too}
In this paper, static or dynamic bindings are inserted verbatim in the source code as an example. A version with fully dynamic loading of extensions parameterized from command-line is published at our \cite{mod-SAIL} github repository. 


To see the effect of modularization and dynamic binding on emulator performance we have performed a series of runs for each of the base, \textbf{M}, and \textbf{V} test-benches on a dedicated \texttt{Xeon(R) w5-2455X} system running at $3.2$Ghz.

\begin{figure}[h]
\begin{subfigure}[h]{\columnwidth}
\hspace{-1em}\includegraphics[width=1.1\columnwidth]{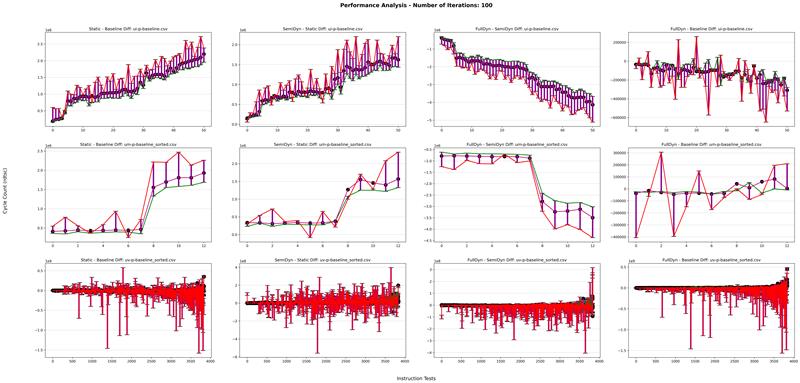}
\caption{Performance of adapted SAIL flow}
\label{fig:perf}
\end{subfigure}

\begin{subfigure}[h]{\columnwidth}
\includegraphics[width=\columnwidth,height=3cm]{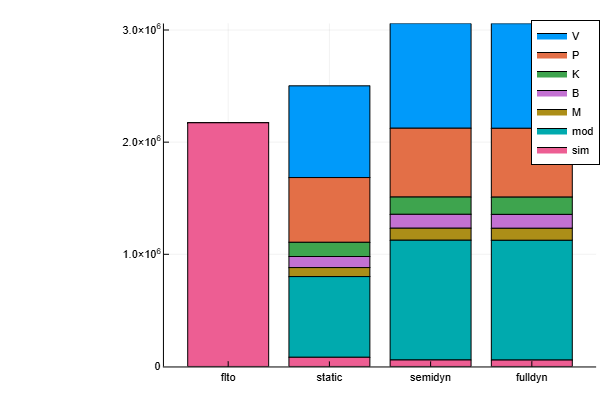}
\caption{Impact of changes on binaries}
\label{fig:sizes}
\end{subfigure}
\caption{Performance study results}
\end{figure}

The impact on binaries (\texttt{text+data+bss}) for the orig, base, static/dynamic is on Fig.\ref{fig:sizes}. We can observe that changes (with \textbf{M}, \textbf{B}, \textbf{K}, \textbf{P} \& \textbf{V}) do not lead to excessive size expansion and that \textbf{LTO} is effective. 

\vspace{-1.5em}
\section{Conclusions and future work}
\vspace{-.5em}

In this paper we have shown that SAIL modularization is a promising path for improving compositionality of RISC-V designs and their ISAs. Addressing this is of paramount importance as systems grow in complexity and variation, and our work makes first steps in this direction for RISC-V ISA emulators. Vendors can now exchange dynamic libraries instead of patches to the golden model, and keep internals concealed (when necessary) while relying on their partners to deliver well-tested, verified and validated extension modules. 


We intend to improve upon the current version of modular SAIL by extending the loader to perform \emph{ondemand} loading of extensions, in addition to provisions for \emph{certifying} used dynamic libraries at load time. 
Extending this work to cover more extensions, primarily \textbf{F} \& \textbf{D} as modules and solving the problem of modular \emph{inheritance} of e.g., floating-point and/or vector features in other extensions such as \textbf{VectorCrypto} would be an interesting followup.
Another promising direction is to apply AI \& ML techniques to automatically instantiate required modules provided a user-friendly description, trained on the published RISC-V specs.
We hope that our findings will be helpful in adapting tools such as SAIL compiler to produce more modular and compositional extension modules. 

\vspace{-1em}
{\footnotesize
\printbibliography 
}

\end{document}